\documentclass[11pt]{article}

\usepackage{amsmath, amsthm, amsfonts}
\usepackage[margin=2.5cm]{geometry}
\usepackage{amssymb}
\usepackage[cp1250]{inputenc}
\usepackage{hyperref}
\hypersetup{
    colorlinks=true,
    linkcolor=blue,
    filecolor=magenta,      
    urlcolor=cyan,
}

\newtheorem{thm}{Theorem}[section]

\theoremstyle{definition}
\newtheorem{defn}[thm]{Definition}
\theoremstyle{remark}

\title{Jordan Form and Quantum Tomography}
\author{Artur Czerwi{\'n}ski\\
\small ResearchGate: \url{www.researchgate.net/profile/Artur_Czerwinski} \\
  \small 1. Institute of Physics\\
  \small Nicolaus Copernicus University\\
  \small 87-100 Toru{\'n}\\
	\small 2. Center for Theoretical Physics \\
\small Polish Academy of Sciences \\
\small 02-668 Warszawa
}

\begin{document}
\maketitle

\abstract{In this brief article we indicate a connection between Jordan normal form of a square matrix and the stroboscopic approach to quantum tomography. We show that the index of cyclicy of a generator of evolution, which receives much attention in the stroboscopic tomography, can be defined and computed by referring to the Jordan decomposition of a square matrix. The result presented in this work shows a relation between terminology from quantum tomography and linear algebra.
}

\section{Introduction}

The term \textit{quantum tomography} one usually assigns to a wide variety of methods which aim to reconstruct an accurate representation of a quantum system (e.g. quantum wavefunction or density matrix) on the basis of data accessible from an experiment. According to one of the most well-known models of quantum state tomography, the so-called static tomography model, one can reconstruct the density matrix of a quantum system only if one can measure $N^2 - 1$ distinct observables (where $N = dim \mathcal{H}$). The major drawback of this model is the fact that the number of distinct observables required for quantum tomography increases quadratically with the dimension of the Hilbert space. Nevertheless, the static model of quantum tomography can be found in many articles, such as \cite{altepeter04,genki03}. When it comes to the quantum wavefunction, there have been many approaches to measure it in an indirect way. Some of them are based on the Gerchberg-Saxton algorithm, which allows to retrieve the phase of the wavefunction from intensity measurements \cite{saxton}. Some recent applications of this algorithm can be found in\cite{czerwin01,czerwin02}. In 2011 a breakthrough paper concerning quantum tomography of wavefunction was published in which a direct method of state reconstruction was presented \cite{bamber}. This paper initiated a new approach to quantum tomography - direct reconstruction of quantum wavefunction or density matrix on the basis of the weak measurement.

We believe that when one is considering problems of quantum tomography, they should also discuss whether a theoretical model can possibly be applied in future experiments. Therefore, in this article we follow the stroboscopic approach to quantum tomography, which focuses on optimal criteria for quantum tomography, i.e. the question we most often ask is: What is the minimal amount of data that allows us to reconstruct the initial density matrix? For the first time the stroboscopic observability appeared in \cite{jam83} and later it was developed in subsequent research papers such as \cite{jam00} and \cite{jam04}. Also a very well-written review paper with all fundamental results has been published \cite{jam12}. Naturally, the notion of the stroboscopic tomography has been the subject of conference presentations. Recently some new ideas have been proposed in \cite{czerwin1501,czerwin1502}.

In the stroboscopic tomography the data for quantum state reconstruction is provided by mean values of some hermitian operators $\{ Q_1, \dots, Q_p\}$, where obviously $Q_i = Q_i^*$. The number of distinct observables is assumed to be lower than in the case of the static approach to quantum tomography (i.e. $p < N^2 - 1$). Therefore, one can conclude that in order to provide data sufficient for state reconstruction each obsrvable has to be measured more than once. In the stroboscopic approach we assume to perform discrete measurements, which seems more natural from experimental point of view. As in this approach we conduct multiple measurement of the same quantity at different time instants, we additionally have to assume that the knowledge about the evolution of quantum system is accessible, e.g. Kossakowski-Lindblad equation is known. Generally, the evolution equation takes the form $\dot{\rho} = \mathbb{L}[\rho]$, where $\mathbb{L}$ is referred to as the generator of evolution.

Although there are many possible aspects concerning this problem, in this article we are mainly interested in the minimal number of distinct observables required for quantum tomography. One can recall the theorem concerning the minimal number of observables \cite{jam00}.

\begin{thm}
For a quantum system with dynamics given by a master equation of the Kossakowski-Lindblad form
\begin{equation}\label{eq:kossakowski}
\dot{\rho} = \mathbb{L} [\rho]
\end{equation}
one can calculate the minimal number of distinct observables for quantum tomography from the formula
\begin{equation}\label{eq:index}
\eta := \max \limits_{\lambda \in \sigma (\mathbb{L})} \{ dim Ker (\mathbb{L} - \lambda \mathbb{I})\},
\end{equation}
where $\sigma(\mathbb{L})$ denotes the spectrum of $\mathbb{L}$, i.e. the set of all eigenvalues of the generator of evolution. This means that for every generator $\mathbb{L}$ there exists a set of observables $\{Q_1, \dots, Q_{\eta}\}$ such that their expectation values at some time instants determine the initial density matrix. Consequently, they also determine the complete trajectory of the state. Apparently, the generator of evolution has to fulfill the condition for completely positive and trace-preserving evolution.
\end{thm}

The number $\eta$ is usually referred to as \textit{the index of cyclicity of a quantum system}. The value of $\eta$ indicates how advantageous it is to employ the stroboscopic tomography instead of the static tomography. The lower the value the more beneficial it is to use stroboscopic rather than static tomography. The index of cycliciy indicates the number of distinct experimental setups that one would have to devise to implement the stroboscopic approach in an experiment. Thus the index of cyclicity receives much attention in quantum tomography and it is the main topic of this article.

\section{Index of cyclicity and Jordan canonical form}

To better understand the mathematical interpretation of the index of cyclicity one can first revise the difference between the algebraic multiplicity and geometric multiplicity. Let us denote the characteristic polynomial of $\mathbb{L}$ by $w(\lambda, \mathbb{L})$. Moreover let assume that the spectrum of $\mathbb{L}$ consists of $r$ distinct eigenvalues, i.e. $\sigma(\mathbb{L}) = \{ \lambda_1, \dots, \lambda_r\}$.

\begin{defn}[Algebraic mutiplicity]
  $n_i$ shall be called the algebraic multiplicity of an eigenvalue $\lambda_i$ if and only if $(\lambda - \lambda_i)^{n_i}$ divides $w(\lambda, \mathbb{L})$ and at the same time $(\lambda - \lambda_i)^{n_i+1}$ does not divide the characteristic polynomial of $\mathbb{L}$.
\end{defn}

\begin{defn}[Geometric multiplicity] $\tilde{n}_i$ shall be called the geometric multiplicity of an eigenvalue $\lambda_i$ if and only if the number of linearly independent eigenvectors corresponding to this eigenvalue is equal $\tilde{n}_i$, which is the same as the dimension of the subspace spanned over the eigenvectors corresponding to $\lambda_i$ (the dimension of the eigenspace) i.e. $dim Ker(\mathbb{L} - \lambda_i \mathbb{I}) = \tilde{n}_i$.
\end{defn}

One can notice that according to the theorem 1. the index of cyclicity of a quantum system with the generator of evolution $\mathbb{L}$ is equivalent to, algebraically speaking, the greatest geometric multiplicity from all eigenvalues.

One can recall the following theorem concerning the Jordan canonical form (Jordan normal form) of a square matrix \cite{matrices}.
\begin{thm}
For any $A \in \mathbb{M}_n (\mathbb{C})$ there exists an invertible matrix $P \in \mathbb{M}_n (\mathbb{C})$ such that
\begin{equation}
P^{-1} A P = J,
\end{equation}
where $J$ is a block diagonal matrix which blocks have the structure
\begin{equation}
J_{p} (\lambda_i) = \left [ \begin{matrix} \lambda_i & 1 & 0 & \dots & 0 \\ 0 & \lambda_i & 1 &\dots & 0 \\ \vdots & \vdots & \ddots &  \ddots & \vdots \\ \vdots & \vdots & \ddots & \lambda_i & 1 \\ 0 & \dots & \dots & 0 & \lambda_i \end{matrix} \right ] \in \mathbb{M}_{p} (\mathbb{C}),
\end{equation}
for some $p \in \mathbb{N}$.
The matrix $J$ shall be called the Jordan form of $A$.
\end{thm}

Jordan canonical form has many interesting properties and two of them will be mentioned here.\\
\textbf{Remark 1.} The total number of blocks that constitute the matrix $J$ is equal to the number of linearly independent eigenvectors of the matrix $A$. \\
\textbf{Remark 2.} The number of blocks in $J$ that correspond to an eigenvalue $\lambda_i$ is equal to the number of linearly independent eigenvectors corresponding to this eigenvalue, i.e. this number is the same as $dim Ker(A - \lambda_i \mathbb{I})$ or, in other words, it is the geometric multiplicity of $\lambda_i$.

As a result of these observations one should understand that the notion of the index of cyclicity of the generator of evolution $\mathbb{L}$ is connected with the Jordan form of $\mathbb{L}$. Namely, in order to determine the index of cyclicity we should be able to find the number of blocks that relate to every eigenvalue and finally choose the greatest figure.

In order to do so for some generator of evolution $\mathbb{L} \in \mathbb{M}_n (\mathbb{C})$ let us define a sequence $q_k (\lambda_i)$ as
\begin{equation}
q_k (\lambda_i) = rank \left( (\mathbb{L} - \lambda_i \mathbb{I})^k \right),
\end{equation}
where $k \in \mathbb{N}$. 

Now we can formulate a theorem 
\begin{thm}
If by $N(m, \lambda_i)$ one denotes the number of blocks of dimension $m$ ($0 \leq m \leq n$) associated with an eigenvalue $\lambda_i$ then $N(m, \lambda_i)$ can be computed from
\begin{equation}
N(m, \lambda_i) = q_{m-1} (\lambda_i) - 2 q_{m} (\lambda_i) + q_{m+1}  (\lambda_i).
\end{equation}
\end{thm}

On the basis of all considerations one can rewrite the formula for the index of cyclicity.
\begin{thm}
The index of cyclicity defined in \eqref{eq:index} for some generator of evolution $\mathbb{L} \in \mathbb{M}_n (\mathbb{C})$ can be computed from
\begin{equation}\label{eq:theorem}
\eta = \max \limits_{\lambda \in \sigma(\mathbb{L})} \left( \sum_{m=1}^n N(m, \lambda) \right).
\end{equation}
\end{thm}

The result \eqref{eq:theorem} indicates the connection between the stroboscopic tomography and the Jordan form of the generator of evolution. 

\section{Summary}

This article presents a connection between the terminology from quantum tomography and linear algebra. Since in general it is relatively difficult to obtain the Jordan form of a generator of evolution of high dimension, therefore, the result introduced in this article should be rather treated as a theoretical observation rather than a result with practical implication. However, if one considers physical systems with generators of evolution for which the Jordan form is known, then the result \eqref{eq:theorem} can simplify the considerations connected with quantum tomography

\section*{Acknowledgement}
Submitted on the Children's Day, dedicated to the memory of the child soldiers who fought and died during the Warsaw Uprising of 1944.

\end{document}